%% file: main.tex
\documentclass{article}
\usepackage{amsthm,amsmath}
\usepackage[utf8]{inputenc} 
\usepackage{amsfonts}
\usepackage{algorithm}
\usepackage{algorithmic}
\usepackage{multirow}
\usepackage[hyphens]{url}
\usepackage{authblk}
\usepackage{graphicx}
\usepackage{tabularx}
\usepackage{color}
\newcommand{\reals}{\mathbb{R}}

\DeclareMathOperator*{\argmax}{arg\,max}
\DeclareMathOperator*{\argmin}{arg\,min}

\newcommand{\todo}[1]{\textcolor{red}{*TODO: #1*}}
\begin{document}
\title{Using Overlapping Methods to Counter Adversaries in Community Detection\thanks{This material is based upon work supported by the United States Air Force under Air Force Contract No. FA8702-15-D-0001 and the Combat Capabilities Development Command Army Research Laboratory (under Cooperative Agreement Number W911NF-13-2-0045). Any opinions, findings, conclusions or recommendations expressed in this material are those of the authors and do not necessarily reflect the views of the United States Air Force or Army Research Laboratory.}}
\date{}

\author[1]{Benjamin~A.~Miller\thanks{Contact author: \texttt{miller.be@northeastern.edu}}}
\author[2]{Kevin~Chan}
\author[1]{Tina~Eliassi-Rad}
\affil[1]{Northeastern University, Boston MA}
\affil[2]{Army Research Laboratory, Adelphi MD}
\maketitle
\begin{abstract}
    When dealing with large graphs, community detection is a useful data triage tool that can identify subsets of the network that a data analyst should investigate. In an adversarial scenario, the graph may be manipulated to avoid scrutiny of certain nodes by the analyst. Robustness to such behavior is an important consideration for data analysts in high-stakes scenarios such as cyber defense and counterterrorism. In this paper, we evaluate the use of overlapping community detection methods in the presence of adversarial attacks aimed at lowering the priority of a specific vertex. We formulate the data analyst's choice as a Stackelberg game in which the analyst chooses a community detection method and the attacker chooses an attack strategy in response. Applying various attacks from the literature to seven real network datasets, we find that, when the attacker has a sufficient budget, overlapping community detection methods outperform non-overlapping methods, often overwhelmingly so. This is the case when the attacker can only add edges that connect to the target and when the capability is added to add edges between neighbors of the target. We also analyze the tradeoff between robustness in the presence of an attack and performance when there is no attack. Our extensible analytic framework enables network data analysts to take these considerations into account and incorporate new attacks and community detection methods as they are developed.

\end{abstract}
\input{intro.tex}
\input{related.tex}
\input{model.tex}
\input{setup.tex}
\input{results.tex}

\input{conclusion.tex}
\section*{Acknowledgements}
The authors wish to thank Christopher L. Smith at MIT Lincoln Laboratory. The idea to consider overlapping community detection methods in this context arose from a conversation between him and the first author.

\bibliographystyle{plain}
\bibliography{bibfile}
\appendix
\input{supplementary.tex}

\end{document}

%% file: intro.tex
\section{Introduction}
\label{sec:intro}
Community detection is an important analytic tool in large graph analysis. When graphs get very large, community detection provides a coarser-grained view of the graph than considering the nodes alone, and can greatly assist network analysts in identifying interesting portions of the graph that require deeper inspection. This makes community detection useful for data triage.

In some applications, we may have data that has been manipulated by  an adversary to point the data analyst in the wrong direction. An infected node in a computer network, for example, may want to avoid being grouped together with other infected nodes, given the risk of having all nodes discovered based on one initial cue to the analyst. A adversary would, in this context, want to spread the relevant nodes across communities so they are not concentrated within any particular highly connected subgraph. This degrades the utility of community detection, rendering it ineffective for the user. 

The adversary's goal becomes more complicated, however, if the analyst applies overlapping community detection~\cite{Xie2013}. In this case, nodes may be assigned to multiple communities, and joining a new community may be insufficient to disassociate from a community that may attract attention. The ability to keep a node within an interesting community has the potential to provide robustness to the analyst when attempting to uncover subnetworks of interest.

In this paper, we evaluate the use of several overlapping community detection methods in the presence of a targeted adversarial attack. We follow the formulation of Kegelmeyer et al.~\cite{Kegelmeyer2018} in which nodes have a measure of interestingness, and the target node wants to avoid being connected to a more interesting community. We formulate the problem as a Stackelberg game in which the defender leads by choosing one of several community detection algorithms and the attacker follows by choosing an attack in response. We find that overlapping community detection methods significantly outperform non-overlapping methods when measured by the target node's position in the ordered list of communities. This is true for even when the adversary's capability is expanded to include the ability to introduce edges between neighbors.

\subsection{Scope and Contributions}
This paper considers the case in which there is a single target node and the adversary's goal is to cause the network analyst to deprioritize it. The attacker is able to add edges between the target node and other nodes to which it is not currently connected. (We also consider a more capable attacker that is able to create edges between pairs of neighbors). As in the work of Kegelmeyer et al.~\cite{Kegelmeyer2018}, each node has a ``temperature'' value denoting its level of interest. Nodes in some cases have attributes, represented by binary vectors. In cases where the nodes come from various classes, we use this information to assign temperatures. The analyst and the attacker both have full visibility of the nodes, edges, and attributes (including temperature), but not their classes. The attacker is able to query the community detection method used by the analyst and obtain the resulting communities, but does not necessarily know the underlying algorithm. The graph consists of a fixed set of nodes, and edges are only manipulated by the attacker, i.e., we are not considering a dynamic graph with a varying topology. The paper's contributions are as follows:
\begin{itemize}
\item We adapt several community detection attacks for use in the current context in which a single vertex wants to evade its initial community, and where the community detection method may yield overlapping communities.
\item We formulate an additional attack that allows the adversary to attempt to create a new community by forming connections between new neighbors (those obtained during the attack) and old ones (neighbors of the target before the attack).
\item We model the defender's optimization as a Stackelberg game with the defender as the leader and the attacker as the follower.
\item We demonstrate on seven real network datasets that overlapping community detection is much more robust to adversarial manipulation than non-overlapping community detection, including the case of the more capable adversary that can introduce connections among neighbors.
\item We analyze the tradeoff between robustness to attack and performance when no attack is present.
\end{itemize}

\subsection{Paper Organization}
The remainder of this paper is organized as follows. Section~\ref{sec:related} briefly summarizes related work on attacks against community detection. Section~\ref{sec:model} provides details on the problem model and formulates the Stackelberg game in which the attacker chooses an attack with respect to the analyst's chosen community detection method. This section summarizes the detection methods and attacks that we consider in our experiments. In Section~\ref{sec:experiments}, we present the results of the Stackelberg game on seven real network datasets, and in Section~\ref{sec:conclusion} we summarize our findings and outline future work.

%% file: related.tex
\section{Related Work}
\label{sec:related}
As attacking machine learning algorithms on graphs has become an active area of research~\cite{Zugner2018,Xu2019,Jin2020b,Mujkanovic2022}, interest in attacking unsupervised learning applications such as community detection has increased. The objectives in the various studies, however, vary considerably. Nagaraja considers a case where an adversary is attempting to de-anonymize a communication network based on community structure, and nodes within the network create new communications to avoid detection~\cite{Nagaraja2010}. A heuristic-based edge rewiring algorithm to hide a specific community is proposed in~\cite{Waniek2018}.  Fionda and Pirr\`{o} introduce the concept of ``safeness''---a measure of how concentrated a node's neighbors are in the node's community---and propose algorithms to reduce the modularity and safeness of a target community by adding and removing edges~\cite{Fionda2017}. Another potentially important aspect is the ``persistence'' of the community, i.e., the extent to which it a community is discovered across multiple attempts at community detection, which is exploited to attack community detection in~\cite{Kumari2021}. Li et al. propose a framework that includes building a generator that is trained to create similar graphs while obscuring the target community~\cite{LiJ2020}.

In one paper, the goal is to destroy the community structure so that the communities identified by various algorithms have low modularity or normalized mutual information, and achieves this using a genetic algorithm~\cite{Chen2019}. Other work considers attacks against a vertex classifier, evaluated with respect to the output of community detection algorithms~\cite{Chen2020b}.

Other work considers individual nodes that do not want to be part of a given community, which is the focus of this paper. Kegelmeyer et al. consider such a scenario, in which a subset of nodes have a ``temperature'' indicating how interesting they are, and the adversary's goal is that the target node not be grouped with interesting (``hot'') nodes~\cite{Kegelmeyer2018}. Chen et al. consider a genetic algorithm that handles the case of target nodes as well as target communities or targeting the overall community structure of the graph, with the fitness function varying depending on the target~\cite{Chen2020a}. While most work has focused on targets that do not want to be included within a community, there is also the issue of identifying non-community members who attempt to join a community~\cite{Jiang2020}.

In addition to attacks against community detection specifically, there has been recent related work focused on attacking graph embeddings~\cite{Bojchevski2019a, Chang2020}. While there are numerous ways to embed vertices into real-valued space---including features based on the nodes' neighborhoods~\cite{Henderson2011} and random walk based methods~\cite{Grover2016}---methods based on random walks in particular have implications for community detection: when there is community structure in a graph, this causes nodes in the same community to be located near each other in the embedded space~\cite{Cavallari2017, Wang2017}. Thus, we consider attacks against embeddings as a potential attack against community detection, as we discuss in Section~\ref{subsec:detection}.

Several papers suggest applying their proposed community detection attacks to overlapping methods, but few have provided any results in this context. One recent paper considers an attack based on a modified degree centrality to make a target node be part of precisely one community when applying an overlapping community detection algorithm~\cite{Liu2022}. We consider this, and other attacks that focus on a single node, in a modified context in which edges can only be added. 

We formulate the analyst's optimization as a Stackelberg game, which is a technique that has been applied in other adversarial graph analysis contexts as well. Recent examples include centrality ranking~\cite{Waniek2021, Waniek2023}, link prediction methods~\cite{Zhou2019}, critical infrastructure defense~\cite{Li2019}, wireless communication~\cite{Li2014}, and privacy preservation~\cite{Zhang2020}. Our present work extends this literature to the area of adversarial community detection.

%% file: model.tex
\section{Problem Model}
\label{sec:model}
We follow the problem model defined in~\cite{Kegelmeyer2018}. We are given a graph $G=(V, E)$, where $V$ is a set of vertices and $E$ is a set of edges. Each vertex has a feature called a ``temperature,''  which we observe as either ``hot,'' ``cold,'' or ``unknown.'' The temperature is quantified by a function $T:V\rightarrow\{-1, 0, 1\}$ such that
\begin{equation*}
T(v)=\begin{cases}
-1 & \mbox{$v$ is cold}\\
 0 & \mbox{$v$'s temperature is unknown}\\
 1 & \mbox{$v$ is hot}
\end{cases}
\end{equation*}
In this scenario, an analyst is determining which vertices in the network require deeper analysis. Rather than analyze all hot vertices (the interesting nodes), which may be overwhelmingly large, the analyst breaks the graph into $k$ communities $C_i\subset V$ such that $\bigcup_{i=1}^kC_i=V$. If communities are disjoint (non-overlapping), then $C_i\cap C_j=\emptyset$ for $i\neq j$. Communities are then ranked according to their average temperature, i.e., community $C_i$'s score is 
\begin{equation}
T(C_i)=\frac{1}{|C_i|}\sum_{v\in C_i}T(v).\label{eq:avgTemp}
\end{equation}
The set of all communities is denoted by $\mathcal{C}=\{C_i | 1\leq i\leq k\}$
The analyst's goal is to find communities that warrant attention, and prioritizes the nodes in hot communities.

We consider the case where there is a vertex attempting to evade the analyst's attention. The vertex is able to create new links (edges), but not delete existing ones. The objective of such an adversarial vertex would be to lower the temperature of the hottest community to which it is assigned. That is, the vertex $v\in V$ adds edges to $E$ with the cost function of minimizing $T(C_i)$ from (\ref{eq:avgTemp}) where $v\in C_i$. We refer to this objective as $T_{\mathrm{comm}}(v)$, the \emph{community temperature} of $v$. If successful, the node will be placed into a cold community and avoid further scrutiny.

It is possible, however, to move to a cold community and still be identified. If the adversary's actions result in \emph{all} communities being relatively cold, for example, being in highest-temperature community would still result in the adversary being found by the analyst. Thus, consider an additional metric that accounts for this possibility. The \emph{rank} of $v$ with respect to the communities $C$, denoted by $r$ and defined as the number of nodes in the union of communities with temperature at least as high as $T_{\mathrm{comm}}(v)$: 
\begin{equation}
    r(v):=\left|\bigcup_{C^\prime\in\{C\in\mathcal{C}|T(C)\geq T_{\mathrm{comm}}(v)\}}{C^\prime}\right|.
\end{equation}
The rank represents the number of nodes the analyst must consider in order to find $v$. The analyst minimizes the rank of an evading target node using a Stackelberg game formulation.
\subsection{Stackelberg Game}
We formulate the counter-adversarial community detection problem as a Stackelberg game, where the analyst is the leader and the attacker is the follower. The move each player makes it to select a technique: The analyst chooses a community detection method, then the attacker chooses an attack. The attacker $v$ will choose whichever attack yields the lowest rank i.e., will choose the attack $E^\prime$, a set of edges that does not exist in the initial graph, to solve
\begin{align}
    \hat{E}_a=&\argmax_{E^\prime\subset\{\{v, u\}|u\in V\}\setminus E}{r_{\mathcal{C}}(v)}\\
    \mathrm{s.t.\ }|E^\prime|\leq & b\\
    G_a= & (V, E\cup E^\prime)\\
    \mathcal{C}= & f_c(G_a),
\end{align}
where  $f_c(\cdot)$ is the analyst's chosen community detection method. Each player is fully aware of the other's capability, and the methods (attacks and defenses) that can be used. The adversary considers each attack strategy listed in Section~\ref{subsec:attacks} and measures the rank for all attack sizes from $0$ to $b$. Of all the attacks, the one that yields the lowest rank is chosen. For a given attack strategy $f_a$, denote this procedure by
\begin{equation}
    E^\prime=f_a(v, G, f_c, b).
\end{equation}
When choosing the best attack for a given target, the adversary maximizes the rank across all strategies, solving
\begin{align}
    \hat{E}^\prime=&\argmax_{f_a\in A}{r_{\mathcal{C}_{f_a}}(v)}\\
    \mathrm{s.t.\ }E^\prime_{f_a} = &f_a(v, G, f_c, b)\\
    G_{f_a}= & (V, E\cup E^\prime_{f_a})\\
    \mathcal{C}_{f_a} = & f_c(G_{f_a}),
\end{align}
where $A$ is the set of attack strategies. The result of this procedure is denoted by
\begin{equation}
    E_{\mathrm{max}}=f_{\mathrm{max}}(v, G, f_c, b, A).
\end{equation}

The analyst's goal is to be as robust to such an attack, defined as \emph{minimizing} the target's rank afterward. The expected rank is the analyst's cost function, and the optimization takes place over all community detection methods available. For each community detection method and each potential target, the analyst evaluates the worst-case rank for each possible attack strategy. Given a set of candidate targets $V_t$, the community detection method selection process is formalized as follows:
\begin{align}
\hat{f}_c=&\argmin_{f_c\in D}\frac{1}{|V_t|}\sum_{v\in V_t}r_{\mathcal{C}_{f_c, v}}(v)\label{eq:defenderCost}\\
\mathrm{s.t.\ }\mathcal{C}_{f_c, v} = &f_c(G^\prime_{f_c, v})\\
G^\prime_{f_c, v} = & (V, E\cup E^\prime_{f_c, v})\\
E^\prime_{f_c, v} = & f_{\mathrm{max}}(v, G, f_c, b, A).
\end{align}
Here $D$ is the set of community detection methods available to the analyst and the objective is the expected rank of the target node, assuming each candidate target is equally likely. Note that the attacker takes the community detection method into account when performing the attack, so the defender must consider all attack--defense pairings when performing the optimization.

\subsection{Community Detection Methods}
\label{subsec:detection}
Within the Stackelberg game, the analyst (leader) considers six community detection methods that may be used by the analyst. The non-overlapping methods make use of the \emph{modularity} metric~\cite{Newman2006}, i.e., 
\begin{equation}
Q:=\frac{1}{2|E|}\sum_{v\in V}\sum_{u\in V}{\left[\mathbb{I}(u\leftrightarrow v)-\frac{1}{2|E|}k_uk_v\right]\mathbb{I}(C(u)=C(v))},\label{eq:modularity}
\end{equation}
where $k_v$ is the degree (number of connections) of vertex $v$, $C(v)$ is the community of $v$ (i.e., $C(v)=i$ if $v\in C_i$), and $\mathbb{I}$ is the indicator function, which resolves to 1 if its argument is true and to 0 otherwise. The notation $u\leftrightarrow v$ is a function that evaluates to true only if $u$ and $v$ share an edge and is false otherwise. Modularity measures the difference between the observed number of edges within and between communities and the expected number of edges if they were randomly rewired. The following community detection methods are used in the experiments.
\begin{itemize}
    \item\emph{Louvain} (LV): A greedy algorithm to maximize modularity (or another quality metric)~\cite{Blondel2008}. Starting with each node in its own community, iteratively move nodes to join communities of their neighbors if it increases partition quality. Once no increase in quality is possible, create a new network where each community from the previous step is a node, and edges from the original graph become multi-edges (or self-loops when the nodes are in the same community). Apply the same procedure to the new graph. Continue until there is no change in partition quality.
    \item\emph{Leiden} (LD): Follow a similar procedure to the Louvain algorithm, but with a ``refinement'' step before aggregation that ensures all communities are well connected~\cite{Traag2019}.
    \item\emph{Clique Percolation} (CP): Create a new graph where each node is a $k$ clique from the original graph. Two nodes share an edge if the corresponding cliques from the original graph share $k-1$ nodes. Communities are defined by the connected components of the new graph. We use the implementation from Reid et al. in our experiments~\cite{Reid2012}.
    \item\emph{Hierarchical Link Clustering} (HLC): For each pair of edges that share a node, compute the edge similarity as the Jaccard coefficient of the neighborhoods of the connected nodes, i.e., the similarity of edges $e_{ik}$ and $e_{jk}$ is $$\frac{|N(i)\cap N(j)|}{|N(i)\cup N(j)|},$$ where $N(i))$ is the neighborhood of node $i$, which includes $i$. Perform hierarchical clustering based on this similarity metric, and communities are determined by the resulting clusters~\cite{Ahn2010}.
    \item\emph{Union of Maximum Spanning Trees Method} (UMST): Compute the union of all maximum spanning trees~\cite{Nocaj2015} using the Jaccard coefficient of the nodes' neighborhoods as edge weights. Create a community around each node consisting of the triangles in the node's neighborhood within the UMST, then merge communities with substantial overlap~\cite{Asmi2020}.
    \item\emph{Neural Overlapping Community Detection} (NOCD): Train a graph neural network (GNN) that outputs the parameters of a Bernoulli--Poisson model~\cite{Zhou2015}, where the probability of an edge existing between nodes $i$ and $j$ is given by $$\Pr(i\leftrightarrow j)=1-\exp\left(-\mathbf{x}_i^\top\mathbf{x}_j\right).$$ Here the vector $\mathbf{x}_i$ is a vector indicating community membership, and is the output of the GNN~\cite{Shchur2019}.
\end{itemize}
The analyst's goal is to choose a method that will perform best in the presence of an adversarial attack, i.e., where the adversary remains among the hottest communities and has relatively small rank.
\subsection{Attacks}
\label{subsec:attacks}
We assume the adversary has a budget $b$ denoting the number of new links that can be created. The adversary may choose any of the following attacks.
\begin{itemize}
\item\emph{Cold and Lonely} (C\&L): First connect to cold nodes, then unknown nodes, then hot nodes. Order nodes in increasing order of degree within a temperature (i.e., connect to nodes with few connections first, many connections later)~\cite{Kegelmeyer2018}.

\item\emph{Stable Structure} (SS): Run community detection several times (which may give different results each time). If two nodes are in the same community every time, they are part of a ``stable structure.'' Find all stable structures and order them in increasing order of temperature. Connect to nodes in each stable structure in this order (random order within a stable structure). Finally, connect to the remaining nodes (those in no stable structure) in increasing order of temperature, breaking ties randomly~\cite{Kegelmeyer2018}. Note that when using an overlapping community detection method, the stable structures may also overlap.
\item\emph{Embedding Attack} (Emb): This attack was developed to attack node embeddings, which can be used to attack community detection~\cite{Bojchevski2019a}. The attack aims to modify the edge set to \emph{maximize} the loss that the node embedding algorithm is trying to minimize, i.e., to solve
\begin{align}
&E^\ast=\argmax_{\hat{E}}{\mathcal{L}(V, \hat{E}, Z^\ast)}\ \ \ Z^\ast=\min_Z{\mathcal{L}(V, \hat{E}, Z)}\label{eq:embed}\\
&\textrm{subject to }|\hat{E}\cup E|-|\hat{E}\cap E|\leq \Delta E,\nonumber
\end{align}
where $Z:V\rightarrow\mathbb{R}^d$ is the $d$-dimensional embedding being learned and $\Delta E$ is the number of edges that can be added or removed by the adversary. The authors use a random-walk-based embedding, where $\|Z(u)-Z(v)\|$ is made smaller the more frequently random walks starting at $u$ reach $v$ or vice versa. We consider a version of this attack where no edges are removed and edges are only added if they connect the target to new neighbors.
\item\emph{Evolutionary Perturbation Attack} (EPA): A genetic algorithm with various modes of operation, attacking overall community detection performance, targeting specific communities for disruption, or targeting a specific node~\cite{Chen2020a}. Like Kegelmeyer et al., the mode in which a specific node is targeted only considers new edges connected to the target. The ``genes'' are sets of edges to add and the fitness function is the ratio of the target's degree before the attack to its degree after the attack. (The fitness is zero if the attack is not successful in moving the target from its initial community.) Genes are selected for subsequent rounds by roulette sampling with probability proportional to their fitness. Genes (attacks) are combined by maintaining their common edges and randomly selecting edges not common to both. Finally, genes  mutate by adding new edges to the attack with probability proportional to the pre-attack distance between their endpoints. The user specifies the number of reproduction rounds and the rate of combination and mutation.
\item\emph{Based Importance Hiding} (BIH): An attack specifically designed for overlapping community detection, in a context different from ours~\cite{Liu2022}. The goal of this method is to take a node that is initially part of several communities and remove it from all but one. It chooses edges to add or remove based on ``degree importance,'' which the authors define with respect to a target node $v$ and a community $C$ as 
$$I(v, C):=\frac{\left(\sum_{u\in N_C^v}\left|N_C^v\cap N_C^u\right|\right)\left(\mathrm{deg}(v)-1\right)}{\mathrm{deg}(v)},$$ where $N_C^v$ is the set of the neighbors of $v$ in community $C$. High-importance edges are added to the community to which the target wants to remain, and removed between the target and communities from which it wants to disassociate. We consider a version of this attack that only adds edges and attempts to connect to join the community with which it has the most non-neighbors (i.e., the most nodes to which a new connection can be established). After connecting to all nodes in the chosen community, the attacker selects another community, continuing until the budget is depleted.
\item\emph{Modularity-Based Attack} (Mod): A baseline method that creates a new edge from the target node to a community that, if the node were to move to that community, would yield the greatest modularity. 
\item\emph{Stable Structure--Introduce Neighbors} (SS-Nbr) We consider one attack that expands the adversary's capability, adding the capacity to create new edges between neighbors. This attack follows the same procedure as SS, but after the target connects to each new neighbor, the new neighbor is also connected to the target's initial set of neighbors (i.e., the neighbors it had in the original graph). This creates the possibility of a new community between the target's new and old neighbors that may dissolve the importance of the target's original community. 
\end{itemize}
Since the attacker moves second, the community detection method is fixed, and the attack optimization takes place with respect to $\hat{f}_c$ in (\ref{eq:defenderCost}). Some strategies (i.e., C\&L, Emb, and Mod) do not consider the specific community detection method while generating perturbations; the attacker only uses the specified method to identify the perturbation that increases the target's rank to the greatest value. The other attacks explicitly use the chosen community detection method when generating perturbations, taking into account the existing community structure presented to the analyst as it determines which edges to add, in addition to determining which perturbation to use after incrementing up to the attack budget.

%% file: setup.tex
\section{Experiments}
\label{sec:experiments}

\subsection{Datasets}
We use seven datasets commonly used in the adversarial graph analysis literature
\begin{itemize}
    \item \emph{CiteSeer}: A network of 3312 scientific publications put into 6 classes based on subject area, with 4732 links representing citations. Each node has a 3703-dimensional binary attribute vector, where each entry represents the presence or absence of a word in the paper.
    \item \emph{Cora}: Another citation network, consisting of 2708 machine learning papers labeled with one of seven categories. The citation network consists of 5429 citations, and each node has a 1433-dimensional binary attribute vector, indicating word presence as with CiteSeer.
    \item\emph{American College Football} (football): A network of 115 nodes representing US college football teams, with 1231 edges indicating which teams played each other during the Fall 2000 season~\cite{Girvan2002}. Each node has a label indicating the conference (out of 12 possible) to which the team belongs.
    \item\emph{Western US Power Grid} (grid): Includes 4941 nodes in the electrical power grid of the western United States, with 6594 edges representing power lines between them~\cite{Watts1998}.
    \item\emph{Network Science Coauthorship} (netsci): A network of 379 network scientists (in the largest connected component) with 914 edges representing coauthorship of articles~\cite{Newman2006}.
    \item\emph{Eu-Email core} (email): An email network of a large European research institution, with 1005 nodes representing users and 25571 directed edges denoting which users emailed others~\cite{Yin2017, Leskovec2007}. Nodes are labeled with one of 42 departments.
    \item\emph{Abu Sayyaf Group} (ASG): A network of the Abu Sayyaf Group, a violent non-state Islamist group operating in the Philippines~\cite{Gerdes2014}. Each node is a member of ASG, and the nodes are linked if the two members both participated in at least one of 105 kidnapping events between 1991 and 2011. The largest connected component in this graph has 207 nodes and 2550 edges.
\end{itemize}
Links to the datasets are available in Appendix~\ref{sec:datasets}.

\subsection{Target Selection and Temperature Assignment}
We select 10 targets from each dataset. To identify these nodes, we compute a stable structure of the graph using 20 trials with the Louvain method. We then consider all stable structures that are \emph{homogeneous}, i.e., all nodes within the structure have the same ground-truth label. For networks without labels (i.e., grid, netsci, and ASG), any stable structure can be used, and the label is taken to be membership in that stable structure. Among the nodes with the same label as the target, temperatures are assigned with probability $\Pr(\mathrm{hot})=0.3$, $\Pr(\mathrm{cold})=0.1$, $\Pr(\mathrm{unknown})=0.6$. For any other label, temperatures are assigned with probabilities of hot and cold reversed.

%% file: results.tex
\subsection{Results}
\label{subsec:results}

\begin{figure}
    \centering
    \includegraphics[width=0.8\textwidth]{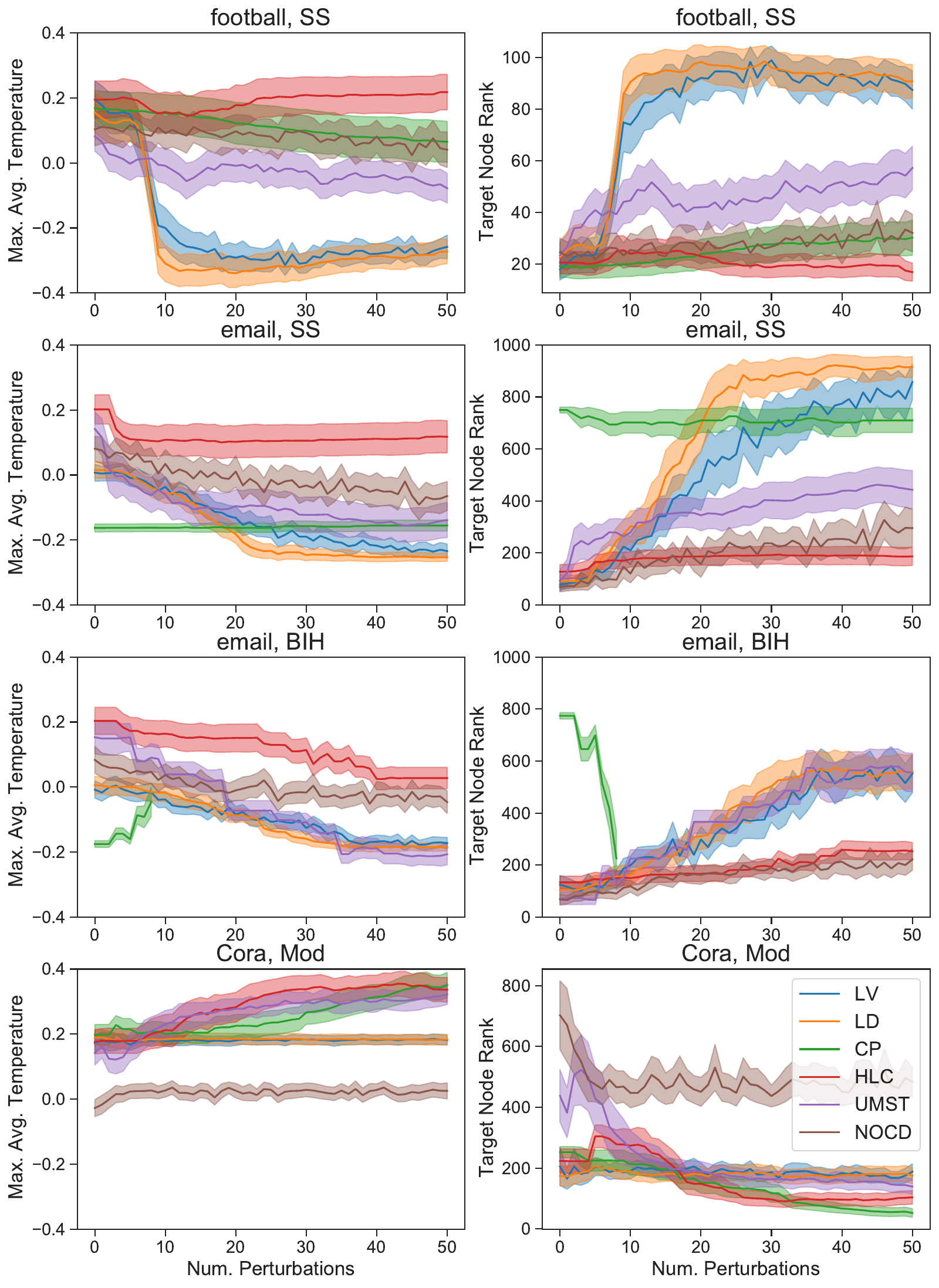}
    \caption{Highlights of attacks applied to various community detection methods. Results are shown in terms of the maximum average temperature across the target node's communities (left column) and the target node's rank in the ordering by community temperature (right column), as the number of edge additions increases from 0 to 50. Higher target node rank is better for the attacker; worse for the analyst. Typical performance is shown in the case of football and email attacked with SS (first and second row, respectively), where we see a substantial change using non-overlapping methods and a more gradual---or even negligible---change using the overlapping methods. Exceptional cases, where the BIH attack is effective against HLC and where the Mod attack does not help the attacker, are shown in the third and fourth rows, respectively. In the typical case, overlapping methods yield less margin for improvement for the attacker, and will be preferred by the analyst.}
    \label{fig:summary_plot}
\end{figure}

We show highlights of the experimental results in Figure~\ref{fig:summary_plot}. The performance of the stable structure attack on the football and email datasets is fairly typical across experiments: The attacks are rather effective against the Louvain and Leiden methods, and somewhat effective against UMST. The attacks are much less effective against CP, with the exception of the email data, where it begins with low temperature and low priority. NOCD, which can use additional side information to make inferences about community structure, also tends to be robust in the face of the attacks. HLC, for the most part, retains a high temperature and a relatively low rank (high priority) for the target's community. Plots for all datasets are included in Appendix~\ref{sec:full_results}.

We observed one exception to HLC's robustness to attack, which is included in the figure: When the attacker uses our variant of BIH, it is effective against HLC on the email data. Investigating this phenomenon further, we noted that BIH tends to break up the target's community and add some of its new neighbors, slightly diluting the concentration of the original community members and reducing the temperature. BIH's focus on the amount of neighborhood overlap seems to work to its advantage in this relatively low-modularity network.

There is also one example in which the attack fails regardless  of the number of perturbations typically yielding a result that is counterproductive to the attacker: the Cora dataset with the modularity-based attack. In this case, the target tends to move to a new community, but one with the same label, and thus the same temperature distribution, which does not improve the target's position. Emb performs much better, likely because nodes from other classes tend to be farther away in the embedding space than those with the same class. While BIH also does not consider temperature, it focuses on lack of overlap between communities rather than modularity maximization, so it is more likely to cause a target to move to a community with a different modal label, and thus a lower temperature.

We present the attacker's average best rank across all dataset/attack/com\-mu\-ni\-ty detection combinations in Table~\ref{tab:all_ranks}. Given the user's community detection method, the attacker chooses the method that minimizes the priority of the target node (i.e., maximizes its rank). Considering the result of the attacker's choice across community detection methods, the user selects the method that results in the highest priority for the target. The top two methods are identified in the table.

\begin{table}
\caption{Average maximum rank across 10 targets for each combination of a dataset, a community detection method (CD), and an attack (third--seventh columns). Standard errors are included, and both averages and standard errors are rounded to the nearest integer. Higher rank is better for the attacker and worse for the data analyst. For each attack strategy, the attacker selects the attack that maximizes the rank, within an attack budget of 50 new edges. Selected strategies are in \textbf{bold}. The community detection methods that yield the smallest and second smallest average target rank are indicated by superscript 1 and 2, respectively, in the second column. Overlapping community detection methods are preferable to the data analyst, often by a large margin.}
\label{tab:all_ranks}
\footnotesize
  \begin{tabularx}{1.04\textwidth}{cccccccc}
    \hline
    dataset& CD   &C\&L&SS&Emb&Mod&BIH\\ \hline
football & LV
	&$49\pm4$
	&$\mathbf{105\pm1}$
	&$46\pm5$
	&$81\pm2$
	&$82\pm2$
\\
football & LD
	&$48\pm4$
	&$\mathbf{105\pm1}$
	&$39\pm4$
	&$70\pm3$
	&$78\pm3$
\\
football & CP$^2$
	&$24\pm3$
	&$\mathbf{33\pm2}$
	&$18\pm1$
	&$23\pm2$
	&$29\pm5$
\\
football & HLC$^1$
	&$23\pm2$
	&$\mathbf{29\pm2}$
	&$20\pm2$
	&$26\pm2$
	&$25\pm2$
\\
football & UMST
	&$53\pm4$
	&$72\pm5$
	&$49\pm6$
	&$65\pm5$
	&$\mathbf{85\pm3}$
\\
football & NOCD
	&$\mathbf{43\pm2}$
	&$\mathbf{43\pm3}$
	&$36\pm2$
	&$31\pm2$
	&$31\pm1$
\\
\hline
netsci & LV
	&$166\pm17$
	&$\mathbf{367\pm5}$
	&$193\pm15$
	&$218\pm7$
	&$232\pm8$
\\
netsci & LD
	&$209\pm16$
	&$\mathbf{361\pm10}$
	&$187\pm22$
	&$224\pm16$
	&$226\pm9$
\\
netsci & CP$^2$
	&$46\pm5$
	&$\mathbf{52\pm6}$
	&$41\pm5$
	&$40\pm6$
	&$37\pm5$
\\
netsci & HLC$^1$
	&$\mathbf{49\pm5}$
	&$41\pm4$
	&$41\pm5$
	&$48\pm5$
	&$46\pm3$
\\
netsci & UMST
	&$87\pm11$
	&$\mathbf{168\pm21}$
	&$111\pm17$
	&$75\pm12$
	&$115\pm17$
\\
netsci & NOCD
	&$\mathbf{172\pm15}$
	&$156\pm13$
	&$87\pm4$
	&$104\pm8$
	&$116\pm8$
\\
\hline
email & LV
	&$\mathbf{986\pm0}$
	&$902\pm22$
	&$249\pm33$
	&$613\pm44$
	&$659\pm18$
\\
email & LD
	&$\mathbf{978\pm5}$
	&$949\pm11$
	&$149\pm24$
	&$642\pm30$
	&$603\pm17$
\\
email & CP
	&$750\pm53$
	&$758\pm46$
	&$742\pm61$
	&$767\pm38$
	&$\mathbf{774\pm30}$
\\
email & HLC$^1$
	&$157\pm19$
	&$197\pm48$
	&$183\pm33$
	&$183\pm24$
	&$\mathbf{284\pm29}$
\\
email & UMST
	&$551\pm73$
	&$521\pm99$
	&$393\pm61$
	&$545\pm31$
	&$\mathbf{607\pm21}$
\\
email & NOCD$^2$
	&$\mathbf{438\pm36}$
	&$403\pm50$
	&$251\pm36$
	&$309\pm29$
	&$284\pm50$
\\
\hline
Cora & LV
	&$\mathbf{2367\pm159}$
	&$2104\pm127$
	&$1988\pm177$
	&$280\pm19$
	&$1974\pm81$
\\
Cora & LD
	&$\mathbf{2600\pm118}$
	&$2129\pm158$
	&$1867\pm173$
	&$229\pm17$
	&$2122\pm42$
\\
Cora & CP$^1$
	&$256\pm118$
	&$\mathbf{277\pm117}$
	&$257\pm118$
	&$252\pm118$
	&$254\pm118$
\\
Cora & HLC$^2$
	&$394\pm102$
	&$\mathbf{426\pm105}$
	&$319\pm80$
	&$308\pm77$
	&$328\pm53$
\\
Cora & UMST
	&$1608\pm220$
	&$\mathbf{2221\pm129}$
	&$1119\pm128$
	&$538\pm112$
	&$1274\pm142$
\\
Cora & NOCD
	&$\mathbf{1233\pm73}$
	&$1221\pm58$
	&$1120\pm91$
	&$795\pm36$
	&$1053\pm44$
\\
\hline
CiteSeer & LV
	&$1116\pm72$
	&$\mathbf{1607\pm33}$
	&$686\pm92$
	&$980\pm74$
	&$1074\pm63$
\\
CiteSeer & LD
	&$1438\pm60$
	&$\mathbf{1603\pm38}$
	&$658\pm89$
	&$990\pm91$
	&$1053\pm83$
\\
CiteSeer & CP$^1$
	&$331\pm80$
	&$326\pm80$
	&$\mathbf{345\pm78}$
	&$314\pm84$
	&$315\pm83$
\\
CiteSeer & HLC$^2$
	&$\mathbf{404\pm58}$
	&$329\pm40$
	&$343\pm27$
	&$314\pm35$
	&$342\pm42$
\\
CiteSeer & UMST
	&$729\pm77$
	&$\mathbf{1328\pm92}$
	&$435\pm41$
	&$566\pm55$
	&$713\pm95$
\\
CiteSeer & NOCD
	&$496\pm27$
	&$667\pm63$
	&$463\pm17$
	&$595\pm41$
	&$\mathbf{807\pm62}$
\\
\hline
grid & LV
	&$1512\pm431$
	&$\mathbf{3678\pm351}$
	&$418\pm116$
	&$2158\pm270$
	&$2545\pm153$
\\
grid & LD
	&$2598\pm336$
	&$\mathbf{4001\pm448}$
	&$497\pm156$
	&$2180\pm251$
	&$2593\pm160$
\\
grid & CP$^2$
	&$\mathbf{448\pm250}$
	&$444\pm250$
	&$444\pm250$
	&$434\pm253$
	&$417\pm254$
\\
grid & HLC$^1$
	&$\mathbf{374\pm26}$
	&$364\pm37$
	&$318\pm34$
	&$324\pm30$
	&$369\pm22$
\\
grid & UMST
	&$2334\pm546$
	&$\mathbf{3160\pm430}$
	&$855\pm87$
	&$1015\pm118$
	&$1172\pm174$
\\
grid & NOCD
	&$993\pm57$
	&$1220\pm104$
	&$892\pm92$
	&$\mathbf{1525\pm138}$
	&$1414\pm147$
\\
\hline
ASG & LV
	&$121\pm15$
	&$\mathbf{188\pm5}$
	&$157\pm5$
	&$154\pm4$
	&$153\pm4$
\\
ASG & LD
	&$129\pm14$
	&$\mathbf{192\pm4}$
	&$149\pm2$
	&$155\pm4$
	&$150\pm2$
\\
ASG & CP$^1$
	&$\mathbf{35\pm3}$
	&$30\pm4$
	&$29\pm2$
	&$32\pm2$
	&$32\pm2$
\\
ASG & HLC$^2$
	&$\mathbf{51\pm12}$
	&$50\pm12$
	&$47\pm9$
	&$44\pm5$
	&$46\pm9$
\\
ASG & UMST
	&$100\pm15$
	&$109\pm13$
	&$123\pm11$
	&$97\pm14$
	&$\mathbf{136\pm8}$
\\
ASG & NOCD
	&$92\pm4$
	&$\mathbf{120\pm7}$
	&$118\pm6$
	&$86\pm6$
	&$114\pm4$
\\
\hline
\end{tabularx}
\end{table}

A few things stand out in the table. First, stable structure is usually the best choice for the attacker, followed by C\&L. Even in cases where the temperature assignment is driven entirely by community structure, these methods that use the temperature information tend to outperform the other methods. Even when restricted to attack strategies that do not consider temperature, however, the overlapping methods almost always outperform LV and LD. (The exception is CP on email, where the attacker's best option is to add no edges, as shown in Figure~\ref{fig:summary_plot}.)

When choosing the community detection method that minimizes the attacker's average rank (i.e., makes it higher priority for the analyst), the two non-overlapping methods always perform worst. HLC is always present in the top two, with the other being either CP or NOCD. UMST is consistently superior to the non-overlapping methods, but typically underperforms with respect to the other overlapping methods. Looking more deeply into the results, we note that UMST is more likely than other overlapping methods to put a node into a single community, which may hinder its performance in this particular task.

We also tested the EPA method. Each gene is an attack (a set of edges to add), and we seed the population with attacks created by other methods. The fitness function used is the rank of the target node after the attack is performed. (The fitness function is computed with respect to the analyst's community detection method.) We use a population of 100 and run for 10 generations. While this frequently results in the best attack, it is typically within one standard error of the second best, and is time consuming to compute. We therefore omit these results for brevity, as similar performance is always possible with one of the less computationally expensive attacks, and including EPA never impacts the defender's selection of a community detection method.

\begin{figure}
    \centering
    \includegraphics[width=0.96\textwidth]{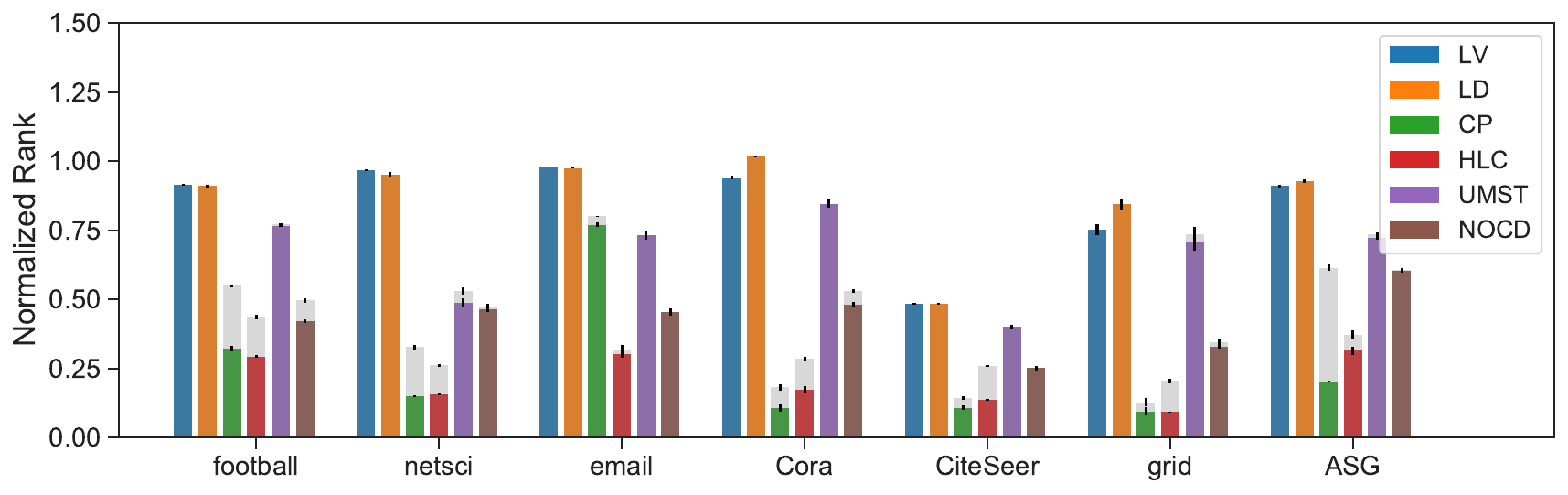}
    \caption{Normalized target rank after attack, where the attacker chooses the strategy that maximizes rank. Bar heights are averages over 10 targets; error bars are standard errors. Higher rank is better for the attacker; the defender will choose the method that yields the lowest rank. Cases where the attacker is given the capability to perform SS-Nbr are shown in grey above the colored bars, which show results when this capability is not available. While the specific method changes depending on capability, in all cases, the defender will choose an overlapping community detection method, and the introduction of the SS-Nbr capability makes NOCD a more attractive option in more cases.}
    \label{fig:stackelberg}
\end{figure}

In the Stackelberg game, the adversary knows the specific target and will select an attack strategy according to that specific node, not the average performance. Average target rank after attack in this scenario is plotted in Figure~\ref{fig:stackelberg}. While the specific aggregated values differ, the top two performers for the analyst remain the same. We also illustrate performance when we expand the adversary's capability and allow SS-Nbr as an attack strategy, keeping the budget at 51 edges. While this strategy usually substantially benefits the attacker when the analyst uses CP or HLC, it has a smaller effect on NOCD, resulting in that method being in the top two for the analyst in additional cases (it overtakes CP in the football and ASG datasets and HLC for CiteSeer). Investigating this matter, we noted that NOCD tends to create relatively few communities (as few as 6 for CiteSeer to as many as 42 for email), while on larger graphs CP and HLC identify hundreds (297 for CP, 739 for HLC). CP and HLC prioritize detecting many small communities, while NOCD identifies fewer larger ones. This propensity to have larger communities may make it more difficult for the target to disassociate from its initial community, despite creating many new triangles with SS-Nbr.

\begin{figure}
    \centering
    \includegraphics[width=0.96\textwidth]{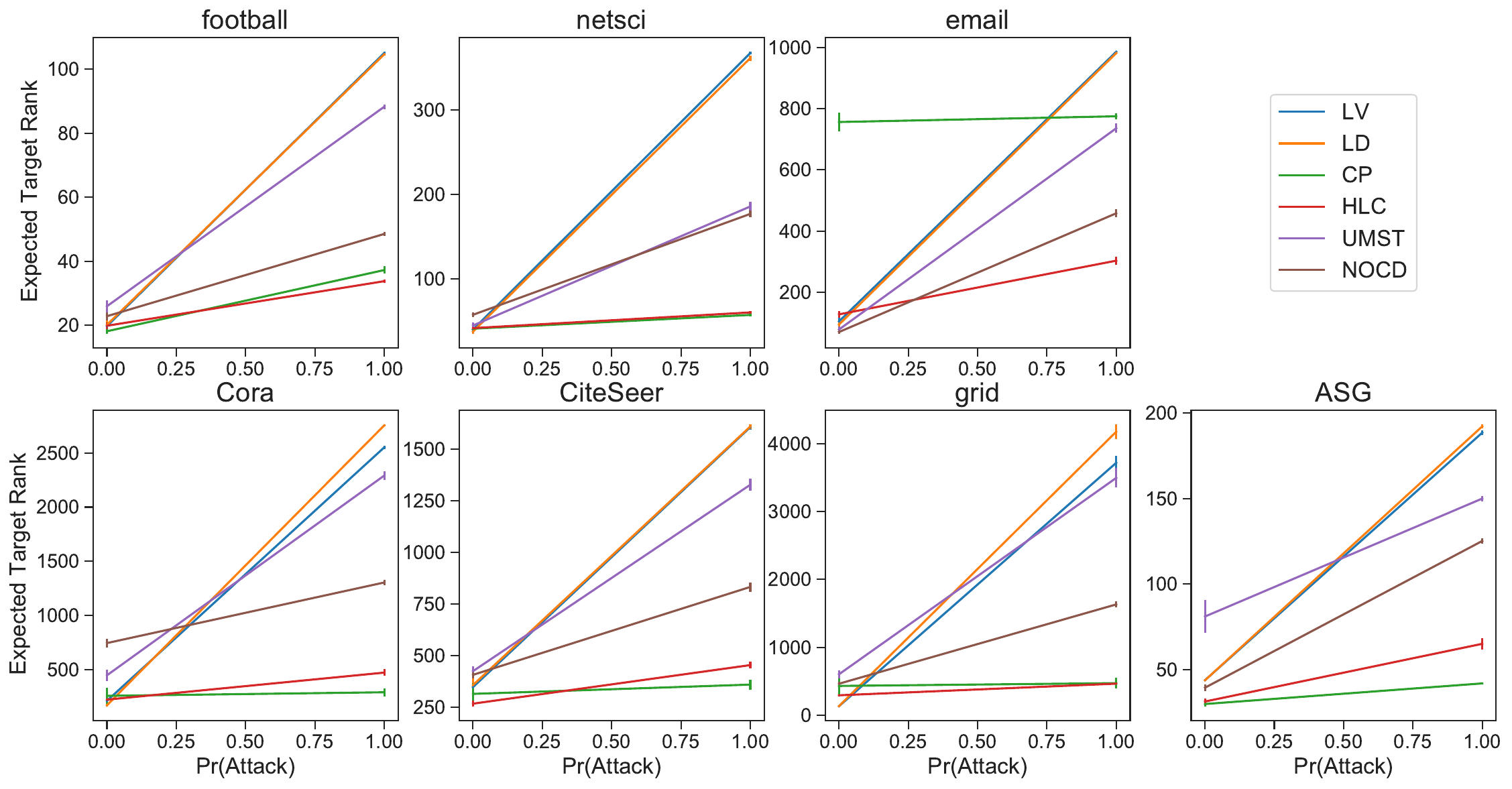}
    \caption{Expected target rank as a function of attack probability. Higher rank is better for the attacker; the defender will choose the method that yields the lowest rank. Non-overlapping methods are often among the best performers for very low attack probabilities, but are quickly overtaken by overlapping methods when the probability of attack increases to greater than 0.1.}
    \label{fig:tradeoff}
\end{figure}

If there is uncertainty regarding whether the target will attack, the analyst must consider this when selecting a community detection method. This alters the defender's optimization formula to be
\begin{align}
f_c=&\argmin_{f_c\in D}\frac{1}{|V_t|}\sum_{v\in V_t}\left[p_A\cdot r_{\mathcal{C}^1_{f_c, v}}(v)+(1-p_A) r_{\mathcal{C}^0_{f_c}}(v)\right]\\
\mathrm{s.t.\ }\mathcal{C}^0_{f_c} = &f_c(G)\\
\mathcal{C}^1_{f_c, v} = &f_c(G^\prime_{f_c, v})\\
G^\prime_{f_c, v} = & (V, E\cup E^\prime_{f_c, v})\\
E^\prime_{f_c, v} = & f_{\mathrm{max}}(v, G, f_c, b, A),
\end{align}
where $p_A$ is the probability of attack. Results taking this consideration into account are shown in Figure~\ref{fig:tradeoff}. The figure includes SS-Nbr as a potential attack strategy. When $p_A=0$, non-overlapping methods perform best in four of seven datasets, but HLC outperforms these methods for any attack probability greater than about 0.057. When there is no attack, HLC's tendency to identify many small communities often elevates smaller hot clusters above those that contain the target, reducing the target's rank at very low probabilities. In some cases, we see a drawback to NOCD's use of fewer communities: at low probability of attack, it often results in lower rank of the target, sometimes substantially so. Its robustness to all attacks considered, however, results in a smaller increase in expected rank than non-overlapping attacks as the probability of attack increases.

%% file: conclusion.tex
\section{Conclusions}
\label{sec:conclusion}
This paper provides an evaluation of overlapping community detection methods as a data triage tool in the presence of adversarial activity. The target node is able to add edges to avoid being placed in a community that will receive greater scrutiny. Since overlapping community detection methods may leave a node in its original community while also placing it in a new one, this has the potential to increase robustness against such an attacker. We formulate the problem as a Stackelberg game in which a data analyst chooses a community detection method and the attacker chooses an attack strategy in response. In our results applying various attacks from the literature to seven real network datasets, we show that overlapping methods do indeed provide a more robust ability to identify the target node, measured by its position in the prioritized list of nodes. This remains the case when the target node is given the capacity to create new connections between its neighbors, though this does improve performance for the attacker. As new attacks and community detection methods are proposed, these can be incorporated into the analytical framework we propose to provide data analysts with the most robust possible community analysis, and a quantification of the tradeoffs between the methods at their disposal.

%% file: supplementary.tex
\section{Dataset Availability}
\label{sec:datasets}
The Cora and CiteSeer datasets used in our experiments were those accompanying the code for the embedding attack~\cite{Bojchevski2019a}. The code is available at \url{https://github.com/abojchevski/node_embedding_attack/tree/master}.

The other datasets are available at the following web locations:
\begin{itemize}
    \item football--- \url{http://websites.umich.edu/~mejn/netdata/football.zip}
    \item netsci---\url{http://websites.umich.edu/~mejn/netdata/netscience.zip}
    \item email---\url{http://snap.stanford.edu/data/email-Eu-core.html}
    \item grid---\url{http://websites.umich.edu/~mejn/netdata/power.zip}
    \item ASG---\url{https://sites.google.com/site/ucinetsoftware/datasets/covert-networks/philippine-kidnappings/}
\end{itemize}

\section{Full Experimental Results}
\label{sec:full_results}
\begin{figure}
    \centering
    \includegraphics[width=.8\textwidth]{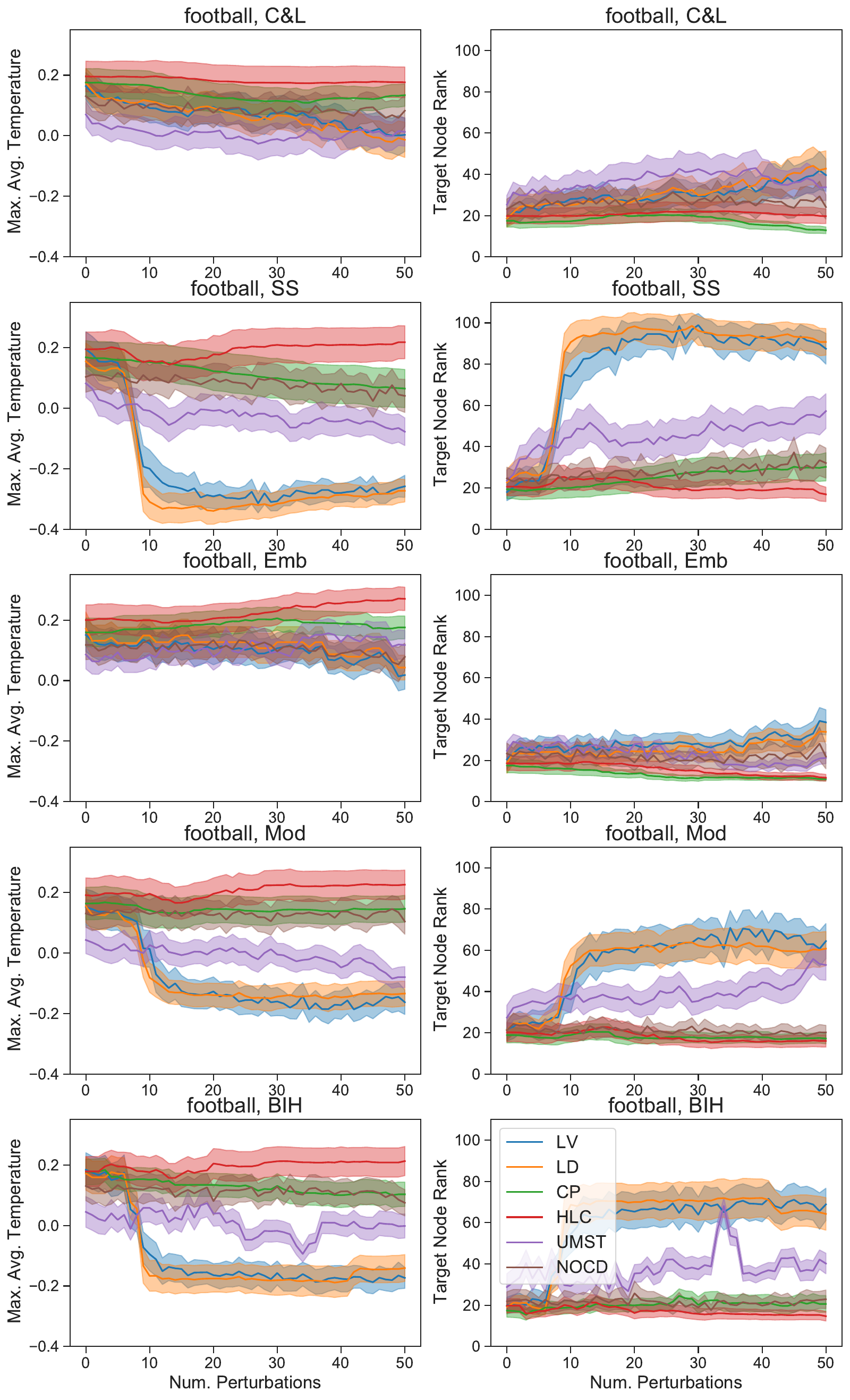}
    \caption{Results of attacks applied to all community detection methods in the college football dataset. Results are shown in terms of the maximum average temperature across the target node's communities (left column) and the target node's rank in the ordering by community temperature (right column), as the number of edge additions increases from 0 to 50. Higher target node rank is better for the attacker; worse for the analyst. HLC provides the best performance (smallest maximum rank) across all detection methods.}
    \label{fig:football}
\end{figure}

\begin{figure}
    \centering
    \includegraphics[width=.8\textwidth]{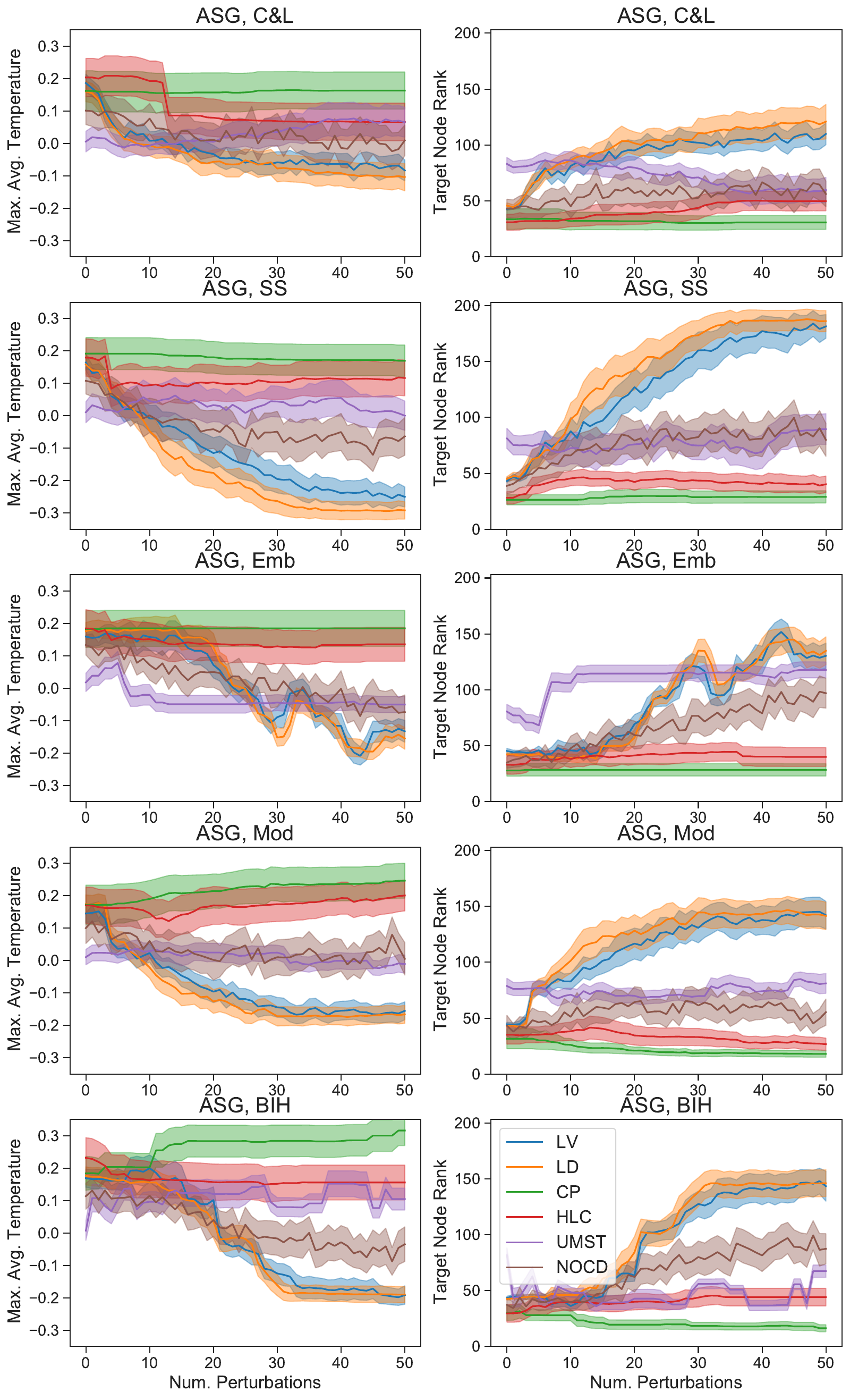}
    \caption{Results of attacks applied to all community detection methods in the Abu Sayyaf Group dataset. Results are shown in terms of the maximum average temperature across the target node's communities (left column) and the target node's rank in the ordering by community temperature (right column), as the number of edge additions increases from 0 to 50. Higher target node rank is better for the attacker; worse for the analyst. CP, HLC, and NOCD all perform well across attacks.}
    \label{fig:kidnap}
\end{figure}

\begin{figure}
    \centering
    \includegraphics[width=.8\textwidth]{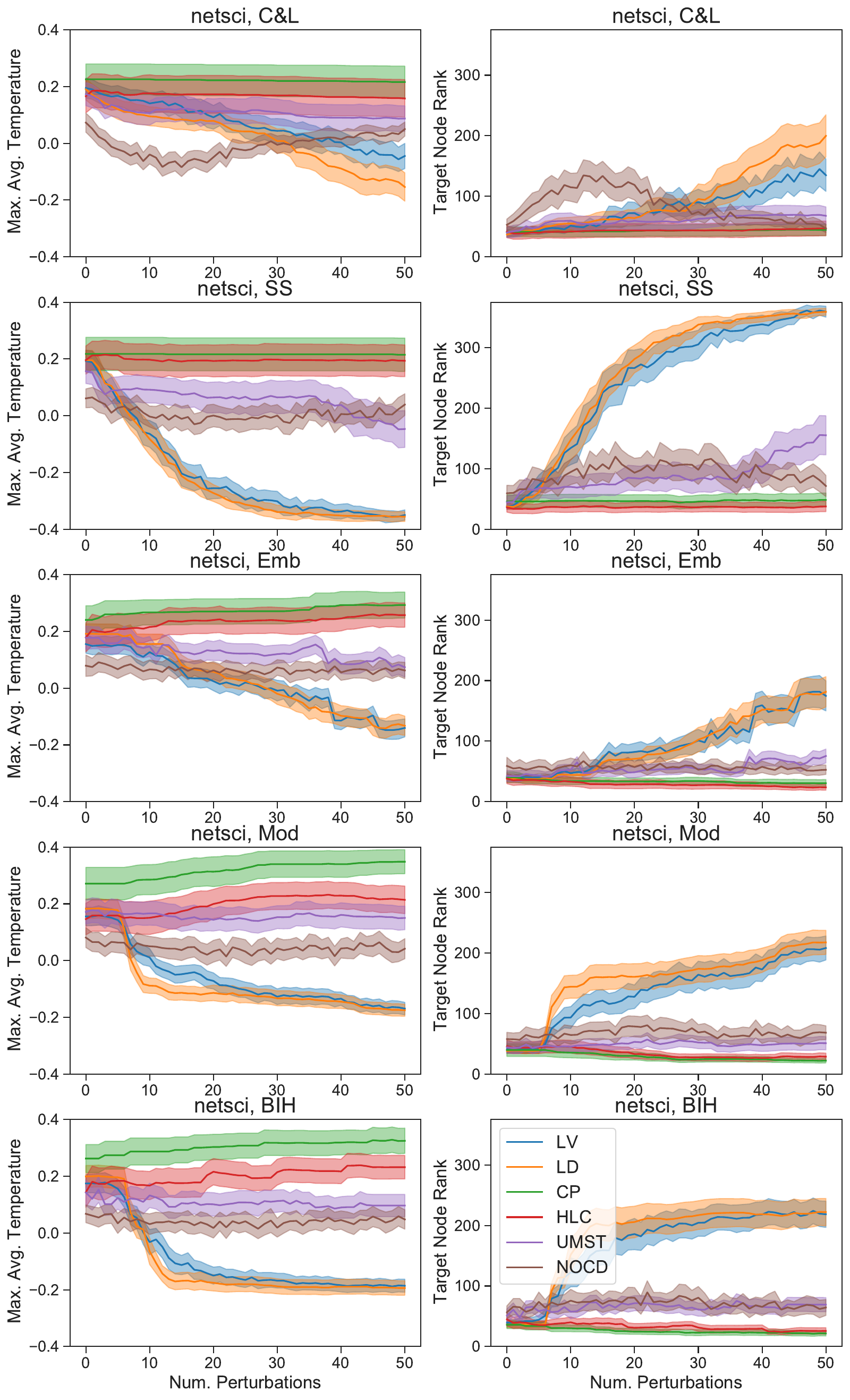}
    \caption{Results of attacks applied to all community detection methods in the network scientist dataset. Results are shown in terms of the maximum average temperature across the target node's communities (left column) and the target node's rank in the ordering by community temperature (right column), as the number of edge additions increases from 0 to 50. Higher target node rank is better for the attacker; worse for the analyst. CP and HLC tend to slightly outperform UMST and NOCD in this case.}
    \label{fig:netsci}
\end{figure}

\begin{figure}
    \centering
    \includegraphics[width=.8\textwidth]{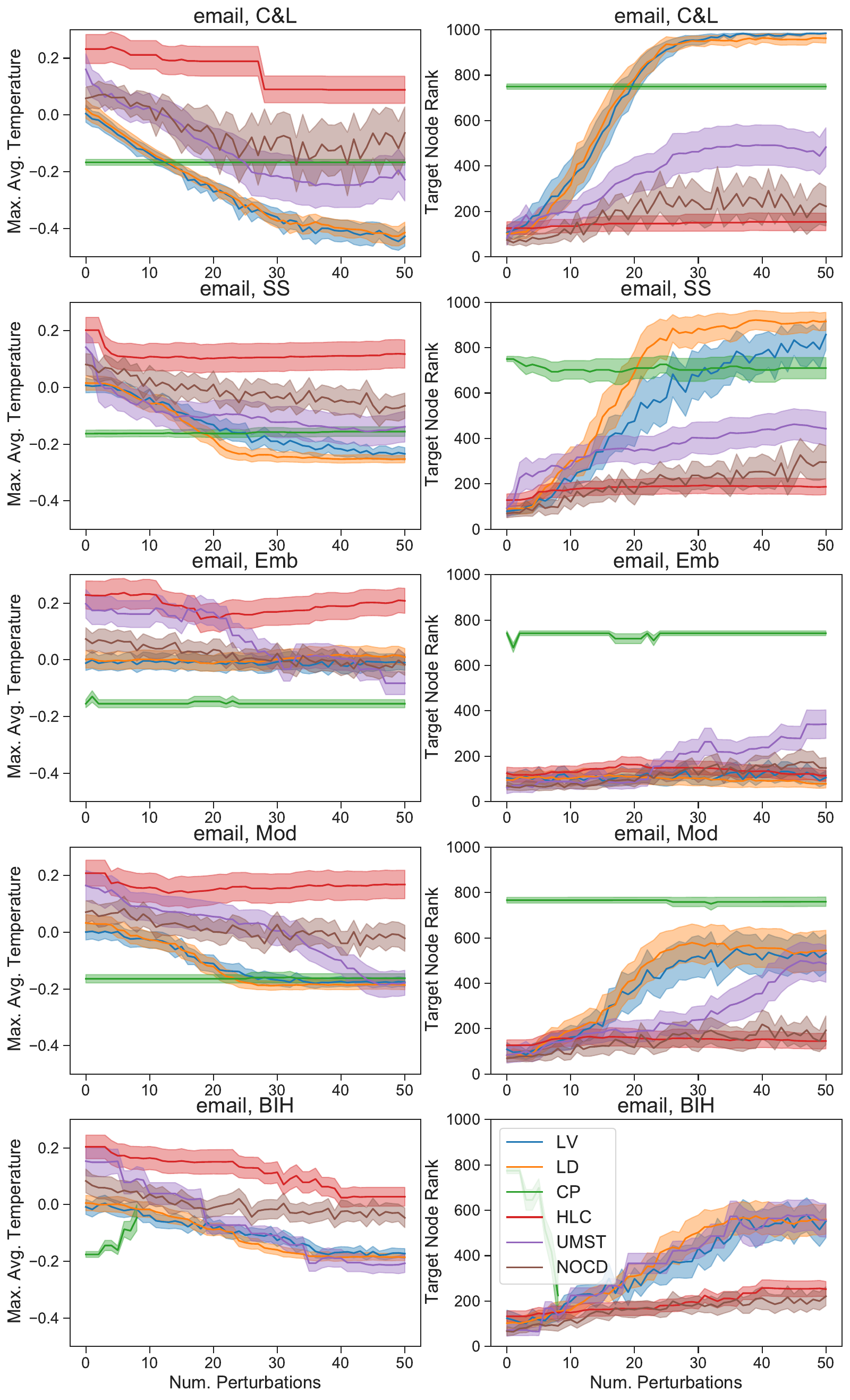}
    \caption{Results of attacks applied to all community detection methods in the European email dataset. Results are shown in terms of the maximum average temperature across the target node's communities (left column) and the target node's rank in the ordering by community temperature (right column), as the number of edge additions increases from 0 to 50. Higher target node rank is better for the attacker; worse for the analyst. CP has particular difficulty with this dataset, typically placing the target outside of any community.}
    \label{fig:email}
\end{figure}

\begin{figure}
    \centering
    \includegraphics[width=.8\textwidth]{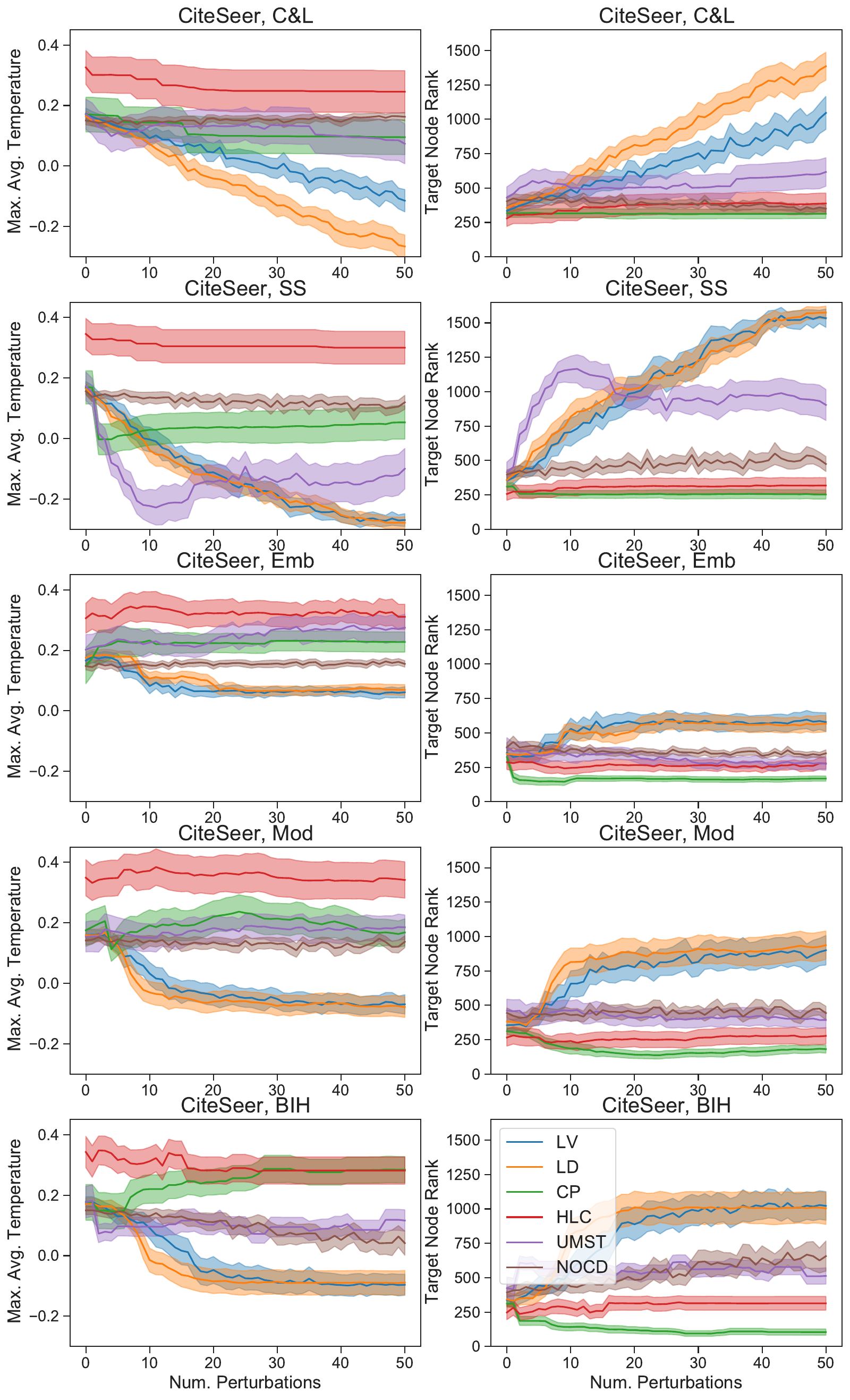}
    \caption{Results of attacks applied to all community detection methods in the CiteSeer citation dataset. Results are shown in terms of the maximum average temperature across the target node's communities (left column) and the target node's rank in the ordering by community temperature (right column), as the number of edge additions increases from 0 to 50. Higher target node rank is better for the attacker; worse for the analyst. For small budgets, UMST is more susceptible to attack than the non-overlapping methods.}
    \label{fig:citeseer}
\end{figure}

\begin{figure}
    \centering
    \includegraphics[width=.8\textwidth]{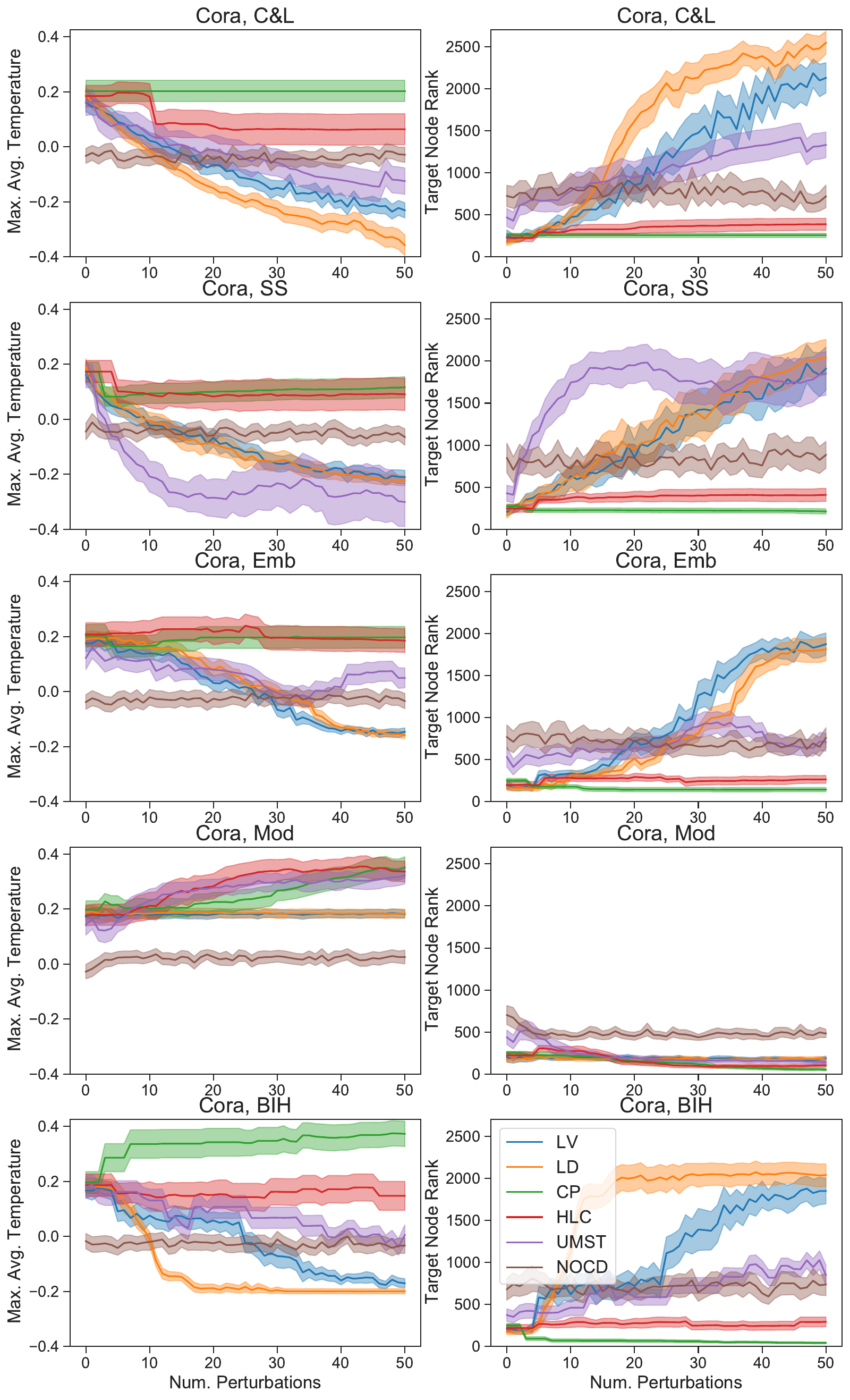}
    \caption{Results of attacks applied to all community detection methods in the Cora citation dataset. Results are shown in terms of the maximum average temperature across the target node's communities (left column) and the target node's rank in the ordering by community temperature (right column), as the number of edge additions increases from 0 to 50. Higher target node rank is better for the attacker; worse for the analyst. Again, UMST is susceptible to attack, and NOCD yields a relatively high rank regardless of attack budget.}
    \label{fig:cora}
\end{figure}

\begin{figure}
    \centering
    \includegraphics[width=.8\textwidth]{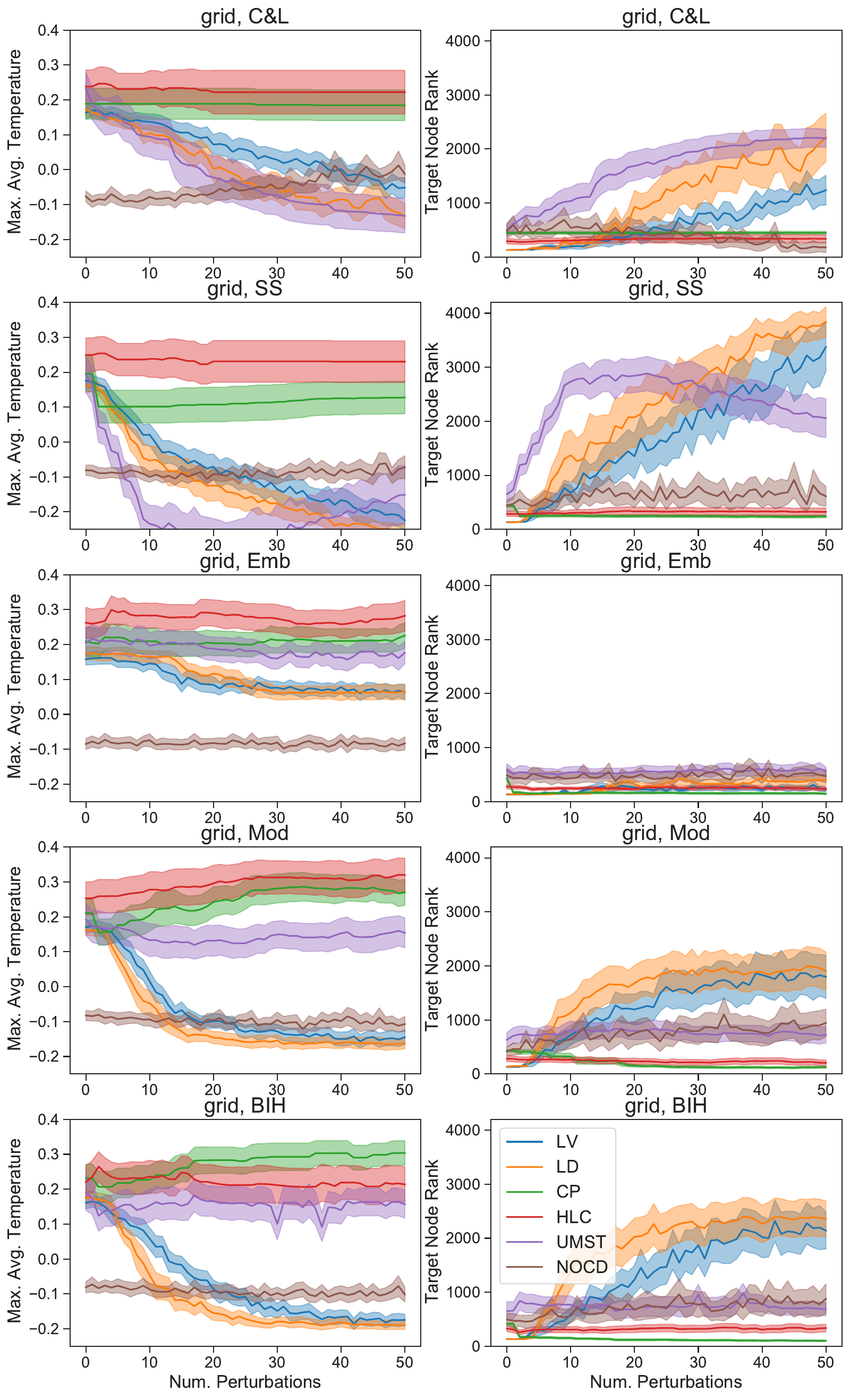}
    \caption{Results of attacks applied to all community detection methods in the power grid dataset. Results are shown in terms of the maximum average temperature across the target node's communities (left column) and the target node's rank in the ordering by community temperature (right column), as the number of edge additions increases from 0 to 50. Higher target node rank is better for the attacker; worse for the analyst. CP and HLC are the top performing methods if there are at least five perturbations.}
    \label{fig:grid}
\end{figure}

Here we plot the full set of experimental results on all datasets: football (Figure~\ref{fig:football}, ASG (Figure~\ref{fig:kidnap}), netsci (Figure~\ref{fig:netsci}), email (Figure~\ref{fig:email}), CiteSeer (Figure~\ref{fig:citeseer}), Cora (Figure~\ref{fig:cora}), and grid (Figure~\ref{fig:grid}). Plots include all attacks except EPA and SS-Nbr.

